\documentclass[nofootinbib,aps,a4paper,superscriptaddress,twocolumn,eqsecnum]{revtex4}

\pdfoutput=1
\usepackage[T1]{fontenc}
\usepackage{amsmath,amssymb,amsfonts,mathtools,bm}
\usepackage{graphicx}
\usepackage{microtype}
\usepackage{xcolor}
\usepackage{url}
\usepackage[linktocpage,colorlinks=true,citecolor=red,linkcolor=blue,urlcolor=magenta,filecolor=magenta]{hyperref}
\allowdisplaybreaks

\newcommand{\mpl}{M_{\rm P}}
\newcommand{\dd}{\mathrm d}
\newcommand{\e}{\mathrm e}
\newcommand{\ii}{\mathrm i}
\newcommand{\R}{\mathcal R}
\newcommand{\Pcal}{\mathcal P}
\newcommand{\avg}[1]{\left\langle #1\right\rangle}
\newcommand{\abs}[1]{\left|#1\right|}
\newcommand{\order}{\mathcal O}
\newcommand{\Var}{\operatorname{Var}}

\begin{document}

\title{Initial conditions for inflation}

\author{Gerasimos Kouniatalis}
\email{gkouniatalis@noa.gr}
\affiliation{National Observatory of Athens, Lofos Nymfon, 11852 Athens, Greece}
\affiliation{Physics Department, National Technical University of Athens, 15780 Zografou Campus, Athens, Greece}

\begin{abstract}
We propose a single-field mechanism that dynamically prepares the initial conditions required for exponential plateau inflation. The inflaton is decomposed into a coarse-grained mode and a high-occupancy, approximately Gaussian band of relativistic nonzero modes. Gaussian--Hartree averaging of the potential
produces an exact variance-dependent minimum, which places the coarse-grained field high on the plateau when the excitation variance is large. Under a translation-invariant microcanonical measure, the available phase-space volume is exponentially concentrated near this minimum, while an order-one homogeneous kinetic-energy fraction is exponentially atypical. During expansion, the preparation modes redshift as radiation, $\sigma^2\propto a^{-2}$ and $\rho_\chi\propto a^{-4}$. Although the induced minimum rapidly moves toward the vacuum, the force acting on the coarse-grained mode is bounded by an exponentially small factor, yielding analytic upper limits on its velocity, displacement, and kinetic fraction. We derive a sufficient condition for the subsequent number of slow-roll e-folds and exhibit a benchmark with more than 60 e-folds. The mechanism also motivates correlated finite-onset signatures: large-scale power suppression with decaying oscillations and a variance-convolution bispectrum.
\end{abstract}


\maketitle

\section{Introduction}
\label{sec:introduction}

Inflation provides a successful framework for generating primordial perturbations and for explaining the observed large-scale homogeneity and spatial flatness of the Universe. Nevertheless, a complete model must also explain why inflation begins: the inflaton must start sufficiently far along its potential, the energy density must be smooth over an expanding region, and the homogeneous kinetic energy must not drive the field through the inflationary domain. These questions are especially sharp for plateau potentials, whose inflationary region is flat but can occupy a restricted part of field space.

The sensitivity of inflation to inhomogeneous field and momentum configurations has been investigated analytically and with numerical relativity~\cite{Linde1985,GoldwirthPiran1990,Goldwirth1991,GoldwirthPiran1992,Brandenberger2017,East2016,JoanaClesse2021,CormanEast2023,Elley2024}. Related approaches include kinetically dominated pre-inflationary phases~\cite{Hergt2019,Contaldi2003}, fluctuation--dissipation dynamics involving additional degrees of freedom~\cite{BasteroGil2016}, and proto-inflationary stages preceding plateau inflation~\cite{DimopoulosArtymowski2017,KouniatalisWaveFunction2025}. Here we consider a more restrictive setup: ordinary Einstein gravity, one canonical scalar field, and no additional matter sector in the preparation mechanism.

The central observation is that the inflaton's own excited nonzero modes modify the coarse-grained potential. For an approximately Gaussian state, exponential moments can be evaluated exactly, and the bare plateau
\begin{equation}
V(\phi)=V_0\left(1-\e^{-\phi/\mu}\right)^2
\end{equation}
develops the Hartree minimum
\begin{equation}
\frac{\varphi_{\rm min}}{\mu}=\frac32\frac{\sigma^2}{\mu^2}.
\end{equation}
A large but finite excitation variance therefore places the preferred coarse-grained field value high on the plateau. Physically, the short-wavelength inflaton excitations act as a transient environment for the long-wavelength mode: they do not add a new field, but they change the force obtained after coarse graining the nonlinear potential. Expansion subsequently removes the same inhomogeneous modes through relativistic redshifting. The induced minimum moves toward the vacuum, but the force available to make the mean field follow it is exponentially small. The release is therefore nonadiabatic from the viewpoint of the coarse-grained mode. This separation of time scales leaves the field high, homogeneous, and slowly moving when ordinary slow roll begins.

We derive the mechanism in closed form, formulate its statistical assumptions explicitly, and obtain analytic bounds on the field displacement and kinetic fraction. We then derive a sufficient condition for more than 60 slow-roll e-folds, give a concrete benchmark, and discuss two linked observational signatures: a finite-onset transfer function and a variance-modulated release bispectrum. The proposal is deliberately falsifiable: Gaussianity, the redshifting law, and the survival of the patch must ultimately be tested with nonequilibrium field theory, lattice simulations, and numerical relativity.

The paper is organized as follows. Section~\ref{sec:model} presents the Einstein--scalar system and the exact Gaussian--Hartree minimum. Section~\ref{sec:preparation} develops the statistical and dynamical preparation mechanism. Section~\ref{sec:inflation-benchmark} derives the e-fold condition and benchmark. Section~\ref{sec:predictions} discusses slow-roll predictions and finite-onset observables, and Sec.~\ref{sec:conclusion} summarizes the results.

\section{Model and exact Hartree mechanism}
\label{sec:model}

\subsection{Dynamical assumptions and notation}

\label{sec:assumptions}

Spacetime obeys ordinary Einstein gravity with reduced Planck mass
\begin{equation}
\mpl\equiv(8\pi G)^{-1/2}.
\end{equation}

There is one real canonical scalar field $\phi$ and no additional matter field in the preparation mechanism.

The nonzero modes form a high-occupancy, approximately Gaussian classical-statistical state over a finite physical momentum band.  The vacuum contribution to $\avg{\chi^2}$ is subtracted; $\sigma^2$ denotes the finite excitation variance. High occupancy is important because it permits the excited sector to be treated as a classical statistical field rather than as a dilute collection of particles~\cite{Berges2004}, while approximate Gaussianity closes the hierarchy of moments at the level required by the Hartree treatment.

Inside one candidate inflationary patch, spatial averages are denoted by $\avg{\cdots}$.  The decomposition is
\begin{equation}
\phi(t,\bm x)=\varphi(t)+\chi(t,\bm x),
\qquad
\avg{\chi}=0,
\qquad
\sigma^2(t)=\avg{\chi^2}.
\label{eq:split}
\end{equation}
The variable $\varphi$ is not an independently introduced degree of freedom. It is the spatially coarse-grained component of the same inflaton field, whereas $\chi$ contains the modes whose wavelengths are shorter than the coarse-graining scale. The preparation mechanism is thus an energy transfer and backreaction effect internal to one scalar field.

A translation-invariant Liouville measure is used when discussing statistical typicality~\cite{RemmenCarroll2013}.

The preparation modes are relativistic and sub-Hubble during the controlled redshifting interval:
\begin{equation}
\frac{k}{a}\gg \max(H,m_{\rm eff}).
\label{eq:subhubble}
\end{equation}
This inequality has two physical roles. It makes the preparation modes insensitive to spacetime curvature over one oscillation and ensures that their group velocity is relativistic. Their amplitude then decreases as $a^{-1}$ and their stress tensor approaches that of radiation, which is the dynamical channel by which the preparation sector disappears.

The dimensionless variables used repeatedly are
\begin{equation}
x\equiv\frac{\varphi}{\mu},
\qquad
s\equiv\frac{\sigma}{\mu},
\qquad
N\equiv\ln a,
\qquad
\epsilon_H\equiv-\frac{\dot H}{H^2}.
\label{eq:defs}
\end{equation}
A dot denotes derivative with respect to cosmic time $t$; a prime on a function of $x$ denotes derivative with respect to $x$; and later a subscript $N$ denotes derivative with respect to the e-fold variable $N$.

\subsection{Einstein--scalar dynamics}

\label{sec:einstein}

The action is
\begin{equation}
S=\int\dd^4x\sqrt{-g}\left[
\frac{\mpl^2}{2}R-\frac12g^{\mu\nu}\partial_\mu\phi\partial_\nu\phi-V(\phi)
\right].
\label{eq:action}
\end{equation}
Variation with respect to $\phi$ gives
\begin{align}
0&=\frac{1}{\sqrt{-g}}\frac{\delta S}{\delta\phi}
=\frac{1}{\sqrt{-g}}\partial_\mu\left(\sqrt{-g}\,g^{\mu\nu}\partial_\nu\phi\right)-V_{,\phi},
\end{align}
therefore
\begin{equation}
\Box\phi-V_{,\phi}=0.
\label{eq:KGcov}
\end{equation}
For
\begin{equation}
\dd s^2=-\dd t^2+a^2(t)\dd\bm x^2,
\qquad H\equiv\frac{\dot a}{a},
\end{equation}
Eq.~\eqref{eq:KGcov} becomes
\begin{equation}
\ddot\phi+3H\dot\phi-\frac{1}{a^2}\nabla^2\phi+V_{,\phi}=0.
\label{eq:KG}
\end{equation}
The stress tensor is
\begin{equation}
T_{\mu\nu}=\partial_\mu\phi\partial_\nu\phi
-g_{\mu\nu}\left[
\frac12g^{\alpha\beta}\partial_\alpha\phi\partial_\beta\phi+V(\phi)
\right].
\end{equation}
Its averaged energy density and isotropic pressure are
\begin{align}
\rho&=\frac12\avg{\dot\phi^2}
+\frac{1}{2a^2}\avg{(\bm\nabla\phi)^2}
+\avg{V(\phi)},
\label{eq:rho}\\
p&=\frac12\avg{\dot\phi^2}
-\frac{1}{6a^2}\avg{(\bm\nabla\phi)^2}
-\avg{V(\phi)}.
\label{eq:pressure}
\end{align}
The spatially averaged Einstein equations are
\begin{align}
3\mpl^2H^2&=\rho,
\label{eq:friedmann}\\
-2\mpl^2\dot H&=\rho+p.
\label{eq:raychaudhuri}
\end{align}
These equations are exact for a homogeneous FRW metric. The gradient term contributes positively to $\rho$ but only one third as much, with the opposite sign, to the isotropic pressure. A relativistic distribution of gradients and time derivatives therefore has $p\simeq\rho/3$, whereas a nearly homogeneous kinetic component has $p\simeq\rho$ and a potential-dominated configuration has $p\simeq-\rho$. These different redshift laws are central to the preparation sequence. When metric inhomogeneities become large, Eqs.~\eqref{eq:friedmann}--\eqref{eq:raychaudhuri} are only the separate-universe approximation; the decisive test must then use numerical relativity.

Averaging Eq.~\eqref{eq:KG} over the patch and using \eqref{eq:split} gives
\begin{equation}
\ddot\varphi+3H\dot\varphi+\avg{V_{,\phi}(\varphi+\chi)}=0.
\label{eq:mean-eom}
\end{equation}
The nonlinear force term will be evaluated exactly within the Gaussian--Hartree approximation~\cite{Boyanovsky1994,RichardSandora2015}. This term, rather than the averaged potential energy alone, determines whether the long mode is driven uphill or downhill. The mechanism therefore depends on the derivative of the coarse-grained effective potential and not merely on an additional contribution to the Friedmann energy density.

\subsection{Exponential plateau potential}

\label{sec:potential}

Take the canonical single-field potential
\begin{equation}
V(\phi)=V_0\left(1-\e^{-\phi/\mu}\right)^2,
\qquad V_0>0,\quad\mu>0.
\label{eq:potential}
\end{equation}
This is an $E$-model-type exponential plateau potential in ordinary Einstein gravity~\cite{Starobinsky1980,FarakosKehagiasRiotto2013,KalloshLindeRoest2013,KehagiasDizgahRiotto2014}. The parameter $V_0$ fixes the inflationary energy scale, while $\mu$ fixes the field-space width over which the potential approaches the plateau. Smaller $\mu/\mpl$ produces a steeper approach to the vacuum but an even flatter asymptotic inflationary region and consequently a smaller tensor-to-scalar ratio at fixed e-fold number~\cite{GengSaridakis2015,KouniatalisSaridakis2025}.

Using $x=\phi/\mu$,
\begin{align}
V&=V_0(1-\e^{-x})^2,
\label{eq:Vx}\\
V_{,\phi}&=\frac{2V_0}{\mu}\e^{-x}(1-\e^{-x}),
\label{eq:Vp}\\
V_{,\phi\phi}&=\frac{2V_0}{\mu^2}\e^{-x}(2\e^{-x}-1).
\label{eq:Vpp}
\end{align}
The unique vacuum is at $x=0$:
\begin{equation}
V(0)=0,
\qquad
V_{,\phi}(0)=0,
\qquad
V_{,\phi\phi}(0)=\frac{2V_0}{\mu^2}>0.
\end{equation}
For $x\gg1$,
\begin{equation}
V=V_0\left[1-2\e^{-x}+\order(\e^{-2x})\right],
\label{eq:plateau-expansion}
\end{equation}
so the positive-$\phi$ side is an exponentially flat plateau. The force $V_{,\phi}$ is proportional to $\e^{-x}$ there, so a field placed at moderately large $x$ can remain nearly stationary for many Hubble times even without an exact symmetry. The initial-condition problem is therefore primarily how to place the field in this region without assigning that value by hand and how to prevent a large initial velocity from carrying it through the plateau.

\subsection{Exact Gaussian--Hartree minimum}

\label{sec:hartree}

\subsubsection{Gaussian exponential moment}

For a Gaussian variable $\chi$ with mean zero and variance $\sigma^2$,
\begin{equation}
P(\chi)=\frac{1}{\sqrt{2\pi\sigma^2}}
\exp\left(-\frac{\chi^2}{2\sigma^2}\right).
\end{equation}
For any real number $q$,
\begin{align}
\avg{\e^{q\chi}}
&=\frac{1}{\sqrt{2\pi\sigma^2}}
\int_{-\infty}^{\infty}\dd\chi
\exp\left(-\frac{\chi^2}{2\sigma^2}+q\chi\right)\nonumber\\
&=\frac{1}{\sqrt{2\pi\sigma^2}}
\int_{-\infty}^{\infty}\dd\chi
\exp\left[-\frac{(\chi-q\sigma^2)^2}{2\sigma^2}
+\frac{q^2\sigma^2}{2}\right]\nonumber\\
&=\exp\left(\frac{q^2\sigma^2}{2}\right).
\label{eq:gaussian-mgf}
\end{align}
Setting $q=-n/\mu$ gives
\begin{equation}
\avg{\e^{-n\chi/\mu}}=\e^{n^2s^2/2}.
\label{eq:exp-average}
\end{equation}
The positive factor on the right-hand side is a direct consequence of the convexity of the exponential. Positive and negative fluctuations do not cancel inside $\e^{-n\chi/\mu}$; rare negative fluctuations receive exponentially larger weight. This is the microscopic origin of the variance-dependent force on the coarse-grained mode.

\subsubsection{Exact averaged potential}

The Gaussian-Hartree potential for the coarse-grained field is
\begin{equation}
V_{\rm H}(\varphi,\sigma)
\equiv\avg{V(\varphi+\chi)}_{\rm G}.
\label{eq:VHdef}
\end{equation}
Using \eqref{eq:potential} and \eqref{eq:exp-average},
\begin{align}
V_{\rm H}
&=V_0\left[
1-2\e^{-\varphi/\mu}\avg{\e^{-\chi/\mu}}
+\e^{-2\varphi/\mu}\avg{\e^{-2\chi/\mu}}
\right]\nonumber\\
&=V_0\left[1-2\e^{-x+s^2/2}+\e^{-2x+2s^2}\right].
\label{eq:VH}
\end{align}
Define $U_{\rm H}\equiv V_{\rm H}/V_0$.  Its first derivative is
\begin{equation}
U_{{\rm H},x}=2\e^{-x+s^2/2}-2\e^{-2x+2s^2}.
\label{eq:UHprime}
\end{equation}
The stationary condition is
\begin{align}
U_{{\rm H},x}=0
&\Longleftrightarrow
\e^{-x+s^2/2}=\e^{-2x+2s^2}\nonumber\\
&\Longleftrightarrow
x_{\rm min}=\frac32s^2.
\label{eq:xtrap}
\end{align}
The second derivative is
\begin{equation}
U_{{\rm H},xx}=-2\e^{-x+s^2/2}+4\e^{-2x+2s^2}.
\end{equation}
At \eqref{eq:xtrap}, both exponentials equal $\e^{-s^2}$, so
\begin{equation}
U_{{\rm H},xx}(x_{\rm min},s)=2\e^{-s^2}>0.
\label{eq:trap-curvature}
\end{equation}
Thus \eqref{eq:xtrap} is a strict minimum. The minimum arises from competition between the first and second exponential moments in the squared plateau potential. The term linear in $\e^{-\phi/\mu}$ and the term quadratic in it are enhanced by different Gaussian factors, so their balance point is shifted away from the vacuum by an amount proportional to the variance. Its energy is
\begin{align}
U_{\rm H}(x_{\rm min},s)
&=1-2\e^{-s^2}+\e^{-s^2}\nonumber\\
&=1-\e^{-s^2}.
\label{eq:trap-energy}
\end{align}
The effective mass of the coarse-grained mode at the minimum is
\begin{equation}
m_{\rm H}^2
\equiv V_{{\rm H},\varphi\varphi}(\varphi_{\rm min},\sigma)
=\frac{2V_0}{\mu^2}\e^{-s^2}.
\label{eq:mH}
\end{equation}
If the Hartree potential dominates the background energy density,
\begin{equation}
H^2\simeq\frac{V_0(1-\e^{-s^2})}{3\mpl^2}.
\end{equation}
Therefore
\begin{equation}
\frac{m_{\rm H}^2}{H^2}
=\frac{6\mpl^2}{\mu^2(\e^{s^2}-1)}.
\label{eq:mass-H-ratio}
\end{equation}
Equations \eqref{eq:xtrap} and \eqref{eq:mass-H-ratio} contain the essential physics:
\begin{equation}
s^2\gg1
\quad\Longrightarrow\quad
x_{\rm min}\gg1,
\qquad
m_{\rm H}^2/H^2\ll1.
\label{eq:high-flat}
\end{equation}
The fluctuations place the preferred mean high on the plateau, while the curvature that could make the mean follow the later motion of that minimum is exponentially weak. In particular, large $s$ simultaneously increases the location of the minimum as $s^2$ and decreases its restoring mass as $\e^{-s^2}$. This opposite scaling is what permits preparation without trapping the field permanently: the fluctuations first select a high value, but the selected long mode becomes too light to adiabatically track the minimum when the variance subsequently redshifts.

\section{Statistical and dynamical preparation}
\label{sec:preparation}

\subsection{Statistical concentration near the high-field minimum}

\label{sec:statistical-high}

Equation \eqref{eq:xtrap} identifies an energy minimum at fixed variance.  To avoid silently assuming that the homogeneous mode was manually placed there, we state the statistical postulate explicitly.

Put the system in a finite physical volume $\mathcal V$ and retain $\mathcal N$ approximately quadratic nonzero-mode quasiparticles.  At fixed coarse-grained field $\varphi$, let the energy available to those modes be
\begin{equation}
E_{\rm bath}(\varphi)=E-\mathcal V V_{\rm H}(\varphi,\sigma)-E_0,
\end{equation}
where $E_0$ is the homogeneous kinetic contribution.  The phase-space volume of $\mathcal N$ classical harmonic oscillators at fixed total bath energy is proportional to
\begin{equation}
\Omega_{\mathcal N}(E_{\rm bath})\propto E_{\rm bath}^{\mathcal N-1}.
\label{eq:phase-volume}
\end{equation}
Consequently, after integrating over the nonzero modes,
\begin{equation}
P(\varphi\mid E,\sigma,E_0)
\propto
\left[E-\mathcal V V_{\rm H}(\varphi,\sigma)-E_0\right]^{\mathcal N-1}
\Theta(E_{\rm bath}).
\label{eq:Pphi}
\end{equation}
Let $\varphi_{\rm min}$ be the minimum \eqref{eq:xtrap}.  The probability ratio is
\begin{align}
\ln\frac{P(\varphi)}{P(\varphi_{\rm min})}
&=(\mathcal N-1)
\ln\left[
1-
\frac{\mathcal V\Delta V_{\rm H}(\varphi)}
{E-\mathcal V V_{\rm H}(\varphi_{\rm min})-E_0}
\right],
\label{eq:Pratio-exact}\\
\Delta V_{\rm H}(\varphi)
&\equiv V_{\rm H}(\varphi)-V_{\rm H}(\varphi_{\rm min})\ge0.
\end{align}
Using $\ln(1-y)\le-y$ for $0\le y<1$,
\begin{equation}
\begin{aligned}
\frac{P(\varphi)}{P(\varphi_{\rm min})}
&\le
\exp\left[-\beta_{\rm mc}\mathcal V\Delta V_{\rm H}(\varphi)\right],\\
\beta_{\rm mc}
&\equiv
\frac{\mathcal N-1}
{E-\mathcal V V_{\rm H}(\varphi_{\rm min})-E_0}.
\end{aligned}
\label{eq:concentration}
\end{equation}
For many modes and a macroscopic patch, the measure is exponentially concentrated near the Hartree minimum. The physical reason is entropic: moving the coarse-grained field away from the minimum stores more energy in one collective coordinate and leaves less energy available to the much larger number of nonzero-mode directions. Even a modest increase in $V_{\rm H}$ is therefore penalized by a power proportional to the number of excited modes. Thus, conditional on the stated measure, the high-field configuration is statistically selected rather than imposed by hand. This is a statement about relative phase-space volume, not about thermal equilibrium or a canonical temperature. If the cosmological measure is not the translation-invariant Liouville measure, or if the state is not approximately ergodic within the relevant macrovariables, this conclusion need not hold.

\subsection{Suppression of a dominant homogeneous kinetic energy}

\label{sec:microcanonical}

Retain $\mathcal N$ approximately quadratic oscillators, including the coarse-grained oscillator $j=0$:
\begin{equation}
E_j=\frac12p_j^2+\frac12\omega_j^2q_j^2.
\end{equation}
The microcanonical measure is
\begin{equation}
\dd\mu_E=C\,\delta\!\left(E-\sum_{j=0}^{\mathcal N-1}E_j\right)
\prod_{j=0}^{\mathcal N-1}\dd q_j\dd p_j.
\label{eq:microcanonical}
\end{equation}
Use
\begin{equation}
q_j=\frac{\sqrt{2E_j}}{\omega_j}\sin\theta_j,
\qquad
p_j=\sqrt{2E_j}\cos\theta_j,
\end{equation}
for which
\begin{equation}
\dd q_j\dd p_j=\frac{1}{\omega_j}\dd E_j\dd\theta_j.
\end{equation}
After the angles are integrated, normalized energy fractions are uniformly distributed on the simplex $E_j\ge0$, $\sum_jE_j=E$.  The simplex volume is
\begin{equation}
\Omega_{\mathcal N}(E)=\frac{E^{\mathcal N-1}}{(\mathcal N-1)!}.
\end{equation}
For
\begin{equation}
y_0\equiv\frac{E_0}{E},
\end{equation}
the marginal density is
\begin{equation}
p(y_0)=(\mathcal N-1)(1-y_0)^{\mathcal N-2},
\qquad0\le y_0\le1.
\label{eq:beta-density}
\end{equation}
Its mean and upper tail are
\begin{align}
\avg{y_0}&=\frac1{\mathcal N},
\label{eq:mean-y0}\\
{\rm Prob}(y_0>f)&=(1-f)^{\mathcal N-1}
\le\e^{-f(\mathcal N-1)}.
\label{eq:tail-y0}
\end{align}
Thus a homogeneous mode carrying an order-one fraction of the total scalar energy is exponentially atypical when no Fourier mode is privileged and $\mathcal N\gg1$. The statement does not require exact equipartition. It follows from the geometry of the constant-energy hypersurface: there are many more directions in phase space in which the energy can be distributed among nonzero modes than there are directions corresponding to a large coherent zero-mode momentum. The exceptional initial data that would overshoot the plateau therefore occupy an exponentially small part of this measure.

For a physical ultraviolet cutoff $\Lambda_{\rm phys}$ in volume $\mathcal V$, the number of independent real modes is approximately
\begin{equation}
\mathcal N\simeq
\frac{\mathcal V}{(2\pi)^3}\frac{4\pi}{3}\Lambda_{\rm phys}^3
=\frac{\mathcal V\Lambda_{\rm phys}^3}{6\pi^2}.
\label{eq:number-modes}
\end{equation}
Equations \eqref{eq:mean-y0}--\eqref{eq:tail-y0} follow exactly from the stated microcanonical measure.

\subsection{Redshift of the preparation modes}

\label{sec:redshift}

For a mode $\chi_{\bm k}$, the linearized equation is
\begin{equation}
\ddot\chi_{\bm k}+3H\dot\chi_{\bm k}
+\left(\frac{k^2}{a^2}+m_{\rm eff}^2\right)\chi_{\bm k}=0.
\label{eq:mode-eom}
\end{equation}
In conformal time $\dd\eta=\dd t/a$, define $u_k=a\chi_k$.  Then
\begin{equation}
u_k''+\left(k^2+a^2m_{\rm eff}^2-\frac{a''}{a}\right)u_k=0.
\label{eq:u-eom}
\end{equation}
Under \eqref{eq:subhubble}, the WKB solution is
\begin{equation}
u_k(\eta)\simeq\frac{C_k}{\sqrt{2k}}\e^{-\ii k\eta},
\qquad
\chi_k=\frac{u_k}{a}\propto a^{-1}.
\label{eq:wkb}
\end{equation}
Therefore
\begin{equation}
\sigma^2\propto a^{-2},\qquad s(N)=s_i\e^{-N}.
\label{eq:s-redshift}
\end{equation}
The variance decays more slowly than the corresponding energy density because it measures field amplitude rather than energy. Each relativistic mode loses one power of $a$ in amplitude, while its physical frequency loses another power of $a$; consequently the quadratic energy contains four inverse powers of the scale factor.
The time-averaged kinetic and gradient energies of a relativistic mode are equal.  Since $\chi_k\propto a^{-1}$ and $k/a\propto a^{-1}$,
\begin{equation}
\rho_{\chi}\simeq\frac12\avg{\dot\chi^2}
+\frac{1}{2a^2}\avg{(\bm\nabla\chi)^2}
\propto a^{-4},
\qquad p_\chi\simeq\frac13\rho_\chi.
\label{eq:rho-rad}
\end{equation}
The Hartree minimum therefore evolves as
\begin{equation}
x_{\rm min}(N)=\frac32s_i^2\e^{-2N}.
\label{eq:moving-min}
\end{equation}
It moves toward the vacuum by an order-one fraction in less than one e-fold. The trajectory $x_{\rm min}\propto\e^{-2N}$ is therefore much faster than ordinary slow-roll motion. The crucial point is that the coarse-grained field is not forced to follow it. Tracking would require a restoring frequency at least comparable to the rate at which the minimum changes, but Eq.~\eqref{eq:mass-H-ratio} shows that the long mode is light precisely when the minimum is high.

\subsection{Freeze-out while the minimum moves}

\label{sec:freeze}

Let
\begin{equation}
S\equiv s_i^2,
\qquad
x_i\equiv\frac32S,
\qquad
q(N)\equiv s^2(N)=S\e^{-2N}.
\label{eq:S-q}
\end{equation}
Evaluate the Hartree force at the initially preferred value $x_i$ while the variance redshifts:
\begin{equation}
U_{{\rm H},x}(x_i,q)
=2\e^{-3S/2+q/2}-2\e^{-3S+2q}.
\label{eq:force-fixed-x}
\end{equation}
At $q=S$ this vanishes exactly.  For $0\le q<S$, it is positive, so the force pushes the field slowly down the plateau.

To find its maximum, differentiate Eq.~\eqref{eq:force-fixed-x} with respect to $q$:
\begin{equation}
\frac{\partial U_{{\rm H},x}}{\partial q}
=\e^{-3S/2+q/2}-4\e^{-3S+2q}.
\end{equation}
The extremum occurs at
\begin{equation}
q_{\rm max}=S-\frac43\ln2,
\label{eq:qmax}
\end{equation}
provided $S\ge(4/3)\ln2$.  At this point the second exponential is one quarter of the first, giving
\begin{equation}
0\le U_{{\rm H},x}(x_i,q)
\le \frac{3}{2^{5/3}}\e^{-S}.
\label{eq:force-max}
\end{equation}
Moreover,
\begin{equation}
U_{\rm H}(x_i,q)
=1-2\e^{-3S/2+q/2}+\e^{-3S+2q}
\ge1-2\e^{-S}.
\label{eq:U-lower}
\end{equation}
During the potential-dominated preparation stage, neglecting $\epsilon_H$ at leading order, the mean equation in e-fold time reduces to
\begin{equation}
x_{NN}+3x_N+F(N)=0,
\qquad
F(N)\equiv\frac{3\mpl^2}{\mu^2}
\frac{U_{{\rm H},x}}{U_{\rm H}}.
\label{eq:xN-eom}
\end{equation}
Equations \eqref{eq:force-max} and \eqref{eq:U-lower} imply
\begin{equation}
0\le F(N)\le F_{\rm max},
\qquad
F_{\rm max}
=\frac{9}{2^{5/3}}\frac{\mpl^2}{\mu^2}
\frac{\e^{-S}}{1-2\e^{-S}}.
\label{eq:Fmax}
\end{equation}
For $x_N(0)=0$, Eq.~\eqref{eq:xN-eom} has the integral solution
\begin{equation}
x_N(N)=-\e^{-3N}\int_0^N\e^{3u}F(u)\dd u.
\label{eq:velocity-integral}
\end{equation}
Hence
\begin{equation}
\abs{x_N(N)}\le\frac{F_{\rm max}}{3}
=\frac{3}{2^{5/3}}\frac{\mpl^2}{\mu^2}
\frac{\e^{-S}}{1-2\e^{-S}}.
\label{eq:velocity-bound}
\end{equation}
After a preparation interval $N_{\rm p}$,
\begin{equation}
\abs{x(N_{\rm p})-x_i}
\le \Delta x_{\rm max}
\equiv
\frac{3N_{\rm p}}{2^{5/3}}\frac{\mpl^2}{\mu^2}
\frac{\e^{-S}}{1-2\e^{-S}}.
\label{eq:displacement-bound}
\end{equation}
This provides the analytic plateau-capture bound: the Hartree minimum itself moves by $\order(x_i)$, but the field moves only by an exponentially small amount $\order(\e^{-s_i^2})$. The phenomenon is analogous to a rapidly removed, very shallow external constraint. The location selected by the constraint changes quickly, yet the constrained coordinate has too little restoring force to respond before Hubble friction erases the induced velocity. The field is consequently released near its original high value rather than transported back to the vacuum.

The physical kinetic fraction follows from $\dot\varphi=\mu Hx_N$:
\begin{equation}
\frac{K_\varphi}{V}
\simeq\frac{\mu^2x_N^2}{6\mpl^2}.
\end{equation}
Using \eqref{eq:velocity-bound},
\begin{equation}
\frac{K_\varphi}{V}
\le
\frac{3}{2^{13/3}}\frac{\mpl^2}{\mu^2}
\left(\frac{\e^{-S}}{1-2\e^{-S}}\right)^2.
\label{eq:kinetic-bound}
\end{equation}
No slow-roll assumption was used to obtain the Gaussian-Hartree minimum; slow variation entered only in replacing $H^2$ by the potential-dominated Friedmann equation when deriving the explicit bounds. This distinction is important: the preparation mechanism is intended to create slow-roll initial data, so its central algebraic result cannot consistently assume slow roll from the outset. The later potential-dominated approximation is used only to convert the already bounded force into conservative displacement and kinetic-energy estimates.

\subsection{Damping of residual homogeneous momentum}

\label{sec:momentum-damping}

While the plateau force is negligible, the mean equation is
\begin{equation}
\ddot\varphi+3H\dot\varphi\simeq0.
\end{equation}
Multiplying by $a^3$ gives
\begin{equation}
\frac{\dd}{\dd t}(a^3\dot\varphi)=0,
\end{equation}
so
\begin{equation}
\dot\varphi\propto a^{-3},
\qquad
K_\varphi=\frac12\dot\varphi^2\propto a^{-6}.
\label{eq:K-a6}
\end{equation}
During $N_{\rm p}$ e-folds,
\begin{equation}
\frac{K_{\varphi,f}}{K_{\varphi,i}}=\e^{-6N_{\rm p}}.
\label{eq:K-suppression}
\end{equation}
For $N_{\rm p}\simeq3$, this factor is $\e^{-18}\simeq1.52\times10^{-8}$.  Thus the statistical suppression of a dominant zero-mode energy in Sec.~\ref{sec:microcanonical} is reinforced dynamically. A homogeneous kinetic component has the stiff equation of state $w=1$ and therefore redshifts faster than radiation and much faster than the plateau energy. Even when the initial zero-mode momentum is not exactly zero, only a few e-folds are needed to make its later effect on the inflationary trajectory negligible.

\subsection{Homogeneity of the prepared patch}

\label{sec:homogeneity}

There are two distinct requirements.

\subsubsection{Suppression of physical gradients}

From Eq.~\eqref{eq:rho-rad}, after $N_{\rm p}$ e-folds,
\begin{equation}
\frac{\rho_{\nabla,f}}{\rho_{\nabla,i}}=\e^{-4N_{\rm p}}.
\label{eq:gradient-suppression}
\end{equation}
Once slow-roll inflation starts, any remaining sub-Hubble gradient contribution continues to be exponentially diluted. The mechanism does not homogenize an arbitrarily large causally disconnected universe. Rather, it requires one initially expanding candidate patch and makes the stress tensor inside that patch progressively more potential dominated. The relevant question for numerical relativity is whether such a patch avoids collapse until this dilution has occurred.

\subsubsection{Uniformity of the local variance}

The Hartree force depends on the locally coarse-grained quantity $\hat\sigma^2$.  Suppose a future inflationary patch contains $\mathcal N_c$ approximately independent correlation cells and the field is Gaussian.  In one cell,
\begin{equation}
\avg{\chi^4}=3\sigma^4,
\end{equation}
so
\begin{equation}
\Var(\chi^2)=\avg{\chi^4}-\avg{\chi^2}^2=2\sigma^4.
\end{equation}
For the sample mean
\begin{equation}
\hat\sigma^2=\frac1{\mathcal N_c}\sum_{a=1}^{\mathcal N_c}\chi_a^2,
\end{equation}
independence gives
\begin{equation}
\Var(\hat\sigma^2)=\frac{2\sigma^4}{\mathcal N_c}.
\end{equation}
Therefore
\begin{equation}
\frac{\sqrt{\Var(\hat\sigma^2)}}{\sigma^2}
=\sqrt{\frac{2}{\mathcal N_c}}.
\label{eq:variance-uniformity}
\end{equation}
Many correlation cells make the variance-induced minimum nearly the same throughout the patch. The local preparation field is therefore self-averaging: different subregions see nearly the same coarse-grained force even though the microscopic field is highly inhomogeneous. This is the statistical part of the homogeneity answer. It does not replace the geometric requirement that at least one sufficiently large, initially expanding patch exists; that requirement must be tested with full numerical relativity.

\subsection{Consistency window for the excitation band}

\label{sec:window}

For a narrow relativistic band with physical root-mean-square momentum $p_i$,
\begin{equation}
\rho_{\chi,i}\simeq p_i^2\sigma_i^2.
\label{eq:rho-p-sigma}
\end{equation}
The band should be sub-Hubble,
\begin{equation}
p_i>H_i,
\label{eq:p-H}
\end{equation}
and its energy should remain below the chosen ultraviolet validity scale $\Lambda_E^4$:
\begin{equation}
p_i^2\sigma_i^2<\Lambda_E^4.
\label{eq:energy-cutoff}
\end{equation}
If the band dominates the initial energy, $3\mpl^2H_i^2\simeq p_i^2\sigma_i^2$.  Equation \eqref{eq:p-H} then requires
\begin{equation}
\sigma_i^2<3\mpl^2.
\label{eq:sigma-window}
\end{equation}
This is why the benchmark below uses $\mu=0.5\mpl$ rather than the larger Starobinsky value: the required dimensionless variance still places the mean high on the plateau, while the physical variance remains inside the simple sub-Hubble window. There is a genuine consistency tension here. Increasing $\sigma_i$ improves high-field preparation, but at fixed physical momentum it also increases the gradient energy and hence $H_i$; beyond Eq.~\eqref{eq:sigma-window}, the same modes can no longer remain parametrically sub-Hubble while dominating the energy density.

For a broad spectrum, define
\begin{equation}
p_{\rm rms}^2\equiv
\frac{\int\dd^3k\,(k/a)^2\abs{\chi_k}^2}
{\int\dd^3k\,\abs{\chi_k}^2}.
\end{equation}
Then Eq.~\eqref{eq:rho-p-sigma} holds with $p_i\to p_{\rm rms}$ up to the usual equality between gradient and kinetic energy for relativistic modes.

\section{Sufficient inflation and an explicit benchmark}
\label{sec:inflation-benchmark}

\subsection{Guaranteed slow-roll duration}

\label{sec:efolds}

For the bare potential \eqref{eq:potential}, the potential slow-roll parameters are
\begin{align}
\epsilon_V
&\equiv\frac{\mpl^2}{2}\left(\frac{V_{,\phi}}{V}\right)^2
=\frac{2\mpl^2}{\mu^2(\e^x-1)^2},
\label{eq:epsilon}\\
\eta_V
&\equiv\mpl^2\frac{V_{,\phi\phi}}{V}
=\frac{2\mpl^2}{\mu^2}
\frac{2-\e^x}{(\e^x-1)^2}.
\label{eq:eta}
\end{align}
Inflation ends at $\epsilon_V(x_{\rm end})=1$, hence
\begin{equation}
x_{\rm end}=\ln\left(1+\frac{\sqrt2\mpl}{\mu}\right).
\label{eq:xend}
\end{equation}
The slow-roll e-fold number from $x$ to $x_{\rm end}$ is
\begin{align}
N_{\rm SR}(x)
&=\frac{1}{\mpl^2}\int_{\phi_{\rm end}}^{\phi}
\frac{V}{V_{,\phi}}\dd\phi\nonumber\\
&=\frac{\mu^2}{2\mpl^2}
\int_{x_{\rm end}}^x(\e^u-1)\dd u\nonumber\\
&=\frac{\mu^2}{2\mpl^2}
\left[
\e^x-x-
\left(\e^{x_{\rm end}}-x_{\rm end}\right)
\right].
\label{eq:Nexact}
\end{align}
At the end of preparation, Eq.~\eqref{eq:displacement-bound} guarantees
\begin{equation}
x_{\rm L}\ge\frac32s_i^2-\Delta x_{\rm max}.
\label{eq:xLbound}
\end{equation}
A sufficient condition for at least $N_{\rm req}$ slow-roll e-folds is therefore
\begin{equation}
N_{\rm SR}\!\left(\frac32s_i^2-\Delta x_{\rm max}\right)
\ge N_{\rm req}.
\label{eq:sufficient-condition}
\end{equation}
In the large-$x$ limit, $N_{\rm SR}\simeq(\mu^2/2\mpl^2)\e^x$, so the leading transparent estimate is
\begin{equation}
s_i^2\gtrsim\frac23
\ln\left(\frac{2\mpl^2N_{\rm req}}{\mu^2}\right),
\label{eq:s-leading}
\end{equation}
with the finite displacement correction included exactly by \eqref{eq:sufficient-condition}. Because the slow-roll duration grows approximately as $\e^x$ while the prepared position grows as $s_i^2$, the required variance depends only logarithmically on the desired number of e-folds. The mechanism therefore does not require an exponentially large field variance to generate an exponentially long plateau trajectory.

\subsection{Benchmark point}

\label{sec:benchmark}

Choose
\begin{equation}
\mu=0.50\mpl,
\qquad
s_i=2.06630895,
\qquad
s_f=0.10.
\label{eq:benchmark-input}
\end{equation}
The initial physical variance and Hartree minimum are
\begin{align}
\sigma_i&=\mu s_i=1.03315\mpl,
\label{eq:sigma-bench}\\
x_i&=\frac32s_i^2=6.40445.
\label{eq:xi-bench}
\end{align}
Thus $\sigma_i^2=1.0674\mpl^2<3\mpl^2$, satisfying Eq.~\eqref{eq:sigma-window}.

The preparation duration needed to reduce $s_i$ to $s_f$ is
\begin{equation}
N_{\rm p}=\ln\frac{s_i}{s_f}=3.02835.
\label{eq:Np-bench}
\end{equation}
The gradient suppression is
\begin{equation}
\e^{-4N_{\rm p}}=5.49\times10^{-6}.
\label{eq:grad-bench}
\end{equation}
The initial Hartree curvature ratio is
\begin{equation}
\frac{m_{\rm H}^2}{H^2}
=\frac{6\mpl^2}{\mu^2(\e^{s_i^2}-1)}
=0.34045.
\label{eq:massratio-bench}
\end{equation}
The analytic displacement and velocity bounds are
\begin{align}
\Delta x_{\rm max}&=0.164708,
\label{eq:dx-bench}\\
\abs{x_N}&\le0.054389.
\label{eq:v-bench}
\end{align}
Therefore
\begin{equation}
x_{\rm L}\ge6.23974.
\label{eq:xL-bench}
\end{equation}
Using Eq.~\eqref{eq:Nexact},
\begin{equation}
N_{\rm SR}(x_{\rm L})\ge63.0000.
\label{eq:N63}
\end{equation}
The kinetic fraction obeys
\begin{equation}
\frac{K_\varphi}{V}
\le1.23\times10^{-4}.
\label{eq:KV-bench}
\end{equation}
The benchmark therefore satisfies the required field-displacement, homogeneity, and kinetic-energy conditions through explicit inequalities. The preparation interval is only about three e-folds, yet the relativistic gradient energy is reduced by nearly six orders of magnitude and a pre-existing homogeneous kinetic component by nearly eight orders of magnitude. At the same time the conservative displacement bound removes only a small fraction of the initial plateau position, leaving enough field range for more than 60 e-folds of conventional slow roll.

\section{Cosmological predictions}
\label{sec:predictions}

\subsection{Slow-roll observables}

\label{sec:observables}

For a canonical single field in the Bunch--Davies state, the leading slow-roll spectra at horizon exit $k=aH$ are standard results~\cite{Mukhanov1985,Sasaki1986,StewartLyth1993}:
\begin{align}
\Pcal_{\R}(k)&=\frac{V_*}{24\pi^2\mpl^4\epsilon_{V*}},
\label{eq:As}\\
\Pcal_T(k)&=\frac{2V_*}{3\pi^2\mpl^4},
\label{eq:PT}\\
r&=16\epsilon_{V*},
\label{eq:r}\\
n_s-1&=-6\epsilon_{V*}+2\eta_{V*}.
\label{eq:ns}
\end{align}
The third slow-roll combination is
\begin{equation}
\xi_V^2\equiv\mpl^4\frac{V_{,\phi}V_{,\phi\phi\phi}}{V^2}
=\frac{4\mpl^4}{\mu^4}
\frac{\e^x-4}{(\e^x-1)^3}.
\label{eq:xi}
\end{equation}
Hence
\begin{align}
\alpha_s
&\equiv\frac{\dd n_s}{\dd\ln k}
=16\epsilon_V\eta_V-24\epsilon_V^2-2\xi_V^2\nonumber\\
&=-\frac{8\mpl^4}{\mu^4}
\frac{\e^x(\e^x+3)}{(\e^x-1)^4}.
\label{eq:alpha}
\end{align}

For $N_*=60$ and $\mu=0.5\mpl$, solving Eq.~\eqref{eq:Nexact} gives
\begin{equation}
x_*=6.191703.
\label{eq:xstar}
\end{equation}
The predictions are
\begin{equation}
\begin{aligned}
 n_s&=0.9670569,\\
 r&=5.38201\times10^{-4},\\
 \alpha_s&=-5.43728\times10^{-4}.
\end{aligned}
\label{eq:predictions}
\end{equation}
Fixing $A_s=2.1\times10^{-9}$ in Eq.~\eqref{eq:As} gives
\begin{align}
\frac{V_0}{\mpl^4}&=1.68009\times10^{-11},
\label{eq:V0bench}\\
\frac{H_*}{\mpl}&=2.36166\times10^{-6},
\label{eq:Hbench}\\
V_*^{1/4}&=2.02250\times10^{-3}\mpl.
\label{eq:Vscale}
\end{align}
For $\mpl=2.435\times10^{18}\,{\rm GeV}$,
\begin{equation}
H_*=5.75\times10^{12}\,{\rm GeV},
\qquad
V_*^{1/4}=4.92\times10^{15}\,{\rm GeV}.
\end{equation}
These observables are those of the bare plateau after the preparation modes have diluted. The preparation mechanism changes the initial phase-space location of the inflaton but does not introduce a new light isocurvature degree of freedom during the observable slow-roll era. Consequently the small value of $r$ is controlled mainly by the plateau width $\mu$, not by an additional conversion mechanism.

\subsection{Comparison with current constraints}

\label{sec:data}

The latest public joint $n_s$--$r$ contours including Planck, SPT, ACT, and BICEP/Keck give~\cite{Balkenhol2025}
\begin{equation}
n_s=0.9682\pm0.0032,
\qquad
r<0.034\quad(95\%\ {\rm confidence}).
\label{eq:latest-cmb}
\end{equation}
A subsequent analysis accepted in July 2026 combines Planck PR4, ACT DR6, and SPT-3G into the CMB-SPA data set and finds~\cite{McDonoughFerreira2026}
\begin{align}
n_s&=0.9693\pm0.0029,
&&\text{CMB-SPA},\label{eq:latest-spa}\\
n_s&=0.9737\pm0.0025,
&&\text{CMB-SPA+DESI DR2}.\label{eq:latest-spa-desi}
\end{align}
For the benchmark in Eq.~\eqref{eq:predictions}, these correspond to
\begin{equation}
0.77\sigma\ \text{below the CMB-SPA central value}
\end{equation}
and
\begin{equation}
2.66\sigma\ \text{below the CMB-SPA+DESI DR2 central value}.
\end{equation}
The benchmark is therefore compatible with current CMB-only constraints but is in greater-than-$2\sigma$ tension with the baseline $\Lambda$CDM CMB-SPA+DESI DR2 inference. The latter shift cannot be described as a clean exclusion of the preparation mechanism: Ref.~\cite{McDonoughFerreira2026} traces it to the correlation of $n_s$ with BAO parameters in the presence of a CMB--DESI tension and explicitly motivates further investigation. Nevertheless, the manuscript cannot claim agreement with all current data without a joint likelihood analysis or a modification that raises the observable-scale tilt.

This comparison tests only the asymptotic slow-roll regime. It establishes that the background trajectory survives CMB-only tests before the preparation-specific signatures are considered. The finite-onset feature derived below is localized at the largest observable scales and requires a dedicated likelihood analysis because cosmic variance, foreground treatment, correlations with standard cosmological parameters, and the oscillatory transfer function cannot be represented by a comparison of central values alone.

\begin{figure}[!t]
\centering
\includegraphics[width=\columnwidth]{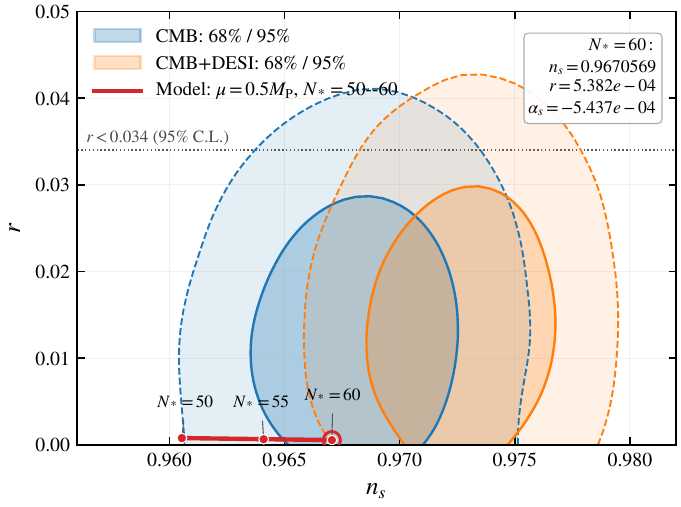}
\caption{Joint $n_s$--$r$ constraints for the asymptotic slow-roll phase. The red trajectory shows the predictions of the exponential plateau with $\mu=0.5\mpl$ as $N_*$ varies from 50 to 60, superimposed on the public 68\% and 95\% CMB and CMB+DESI BAO confidence contours~\cite{Balkenhol2025}. The open circle marks the $N_*=60$ benchmark, and the dotted horizontal line shows the current 95\% upper limit $r<0.034$.}
\label{fig:ns-r-current}
\end{figure}

Figure~\ref{fig:ns-r-current} provides the direct consistency test of the background slow-roll prediction. Across the phenomenologically relevant interval $50\leq N_*\leq60$, the trajectory remains in the small-$r$ region and passes through the displayed joint confidence domains. The benchmark value $r=5.38201\times10^{-4}$ is far below the present tensor bound, so the scalar tilt rather than the tensor amplitude is the more discriminating standard observable for this realization~\cite{ACTExtended,Oikonomou2026}. The plot tests the post-preparation slow-roll phase only; the distinctive finite-onset signatures are examined separately below.

\subsection{Finite-onset transfer function}

\label{sec:transfer}

A near-minimal realization has only a few e-folds between the end of preparation and the exit of the pivot mode.  Define
\begin{equation}
k_{\rm L}\equiv a_{\rm L}H_{\rm L},
\label{eq:kLdef}
\end{equation}
where $a_{\rm L}$ is the scale factor at the end of the variance-dominated preparation stage.

To obtain a closed analytic template, impose the controlled sudden-onset approximation:

after $\eta_{\rm L}$, $H$ and the slow-roll parameters are approximately constant during matching;

$z''/z\simeq2/\eta^2$ for the Mukhanov variable $v_k=z\R_k$;

$v_k$ is in the instantaneous positive-frequency state at $\eta_{\rm L}$~\cite{Danielsson2002,CollinsHolman2005}.

The mode equation is
\begin{equation}
v_k''+\left(k^2-\frac{2}{\eta^2}\right)v_k=0.
\label{eq:MS}
\end{equation}
Write
\begin{equation}
v_k=\frac1{\sqrt{2k}}
\left[
\alpha_k\left(1-\frac{\ii}{k\eta}\right)\e^{-\ii k\eta}
+\beta_k\left(1+\frac{\ii}{k\eta}\right)\e^{\ii k\eta}
\right].
\label{eq:bogoliubov}
\end{equation}
At
\begin{equation}
\eta_{\rm L}=-\frac1{k_{\rm L}},
\qquad
z\equiv-k\eta_{\rm L}=\frac{k}{k_{\rm L}},
\end{equation}
impose
\begin{equation}
v_k(\eta_{\rm L})=\frac{\e^{-\ii k\eta_{\rm L}}}{\sqrt{2k}},
\qquad
v_k'(\eta_{\rm L})=-\ii k v_k(\eta_{\rm L}).
\end{equation}
Solving the two linear equations gives
\begin{equation}
\alpha_k=1-\frac{\ii}{z}-\frac1{2z^2},
\qquad
\beta_k=-\frac{\e^{2\ii z}}{2z^2}.
\label{eq:alpha-beta}
\end{equation}
At late times the growing curvature amplitude is proportional to $\alpha_k-\beta_k$. The nonzero $\beta_k$ represents negative-frequency admixture generated by imposing the vacuum at a finite time rather than in the asymptotic past~\cite{CollinsHolman2005,AgarwalHolmanTolleyLin2013}. Modes with $k\gg k_{\rm L}$ oscillate many times before horizon exit and recover the standard adiabatic result, whereas modes with $k\lesssim k_{\rm L}$ retain information about the onset surface. Therefore
\begin{equation}
\Pcal_{\R}(k)=\Pcal_{\R}^{\rm SR}(k)T^2(k),
\qquad
T^2=\abs{\alpha_k-\beta_k}^2.
\end{equation}
Direct multiplication yields
\begin{equation}
T^2(z)=1+\frac{\cos(2z)}{z^2}
-\frac{\sin(2z)}{z^3}
+\frac{1-\cos(2z)}{2z^4}.
\label{eq:transfer}
\end{equation}
Its limiting forms are
\begin{align}
T^2(z)&=\frac49z^2-\frac1{15}z^4+\order(z^6),
&&z\ll1,
\label{eq:Tsmall}\\
T^2(z)&=1+\order(z^{-2}),
&&z\gg1.
\label{eq:Tlarge}
\end{align}
Thus the simplest observable template is
\begin{equation}
\begin{gathered}
\Pcal_{\R}(k)\propto k^2\quad (k\ll k_{\rm L}),\\
\text{with decaying oscillations around }k_{\rm L}.
\end{gathered}
\label{eq:signature1}
\end{equation}

For the benchmark, the conservative bound leaves approximately
\begin{equation}
N_{\rm L}-N_*=63-60=3
\end{equation}
e-folds between release and pivot exit.  Taking $k_*=0.05\,{\rm Mpc}^{-1}$ and $H_{\rm L}\simeq H_*$,
\begin{equation}
k_{\rm L}\simeq k_*\e^{-3}
=2.49\times10^{-3}\,{\rm Mpc}^{-1}.
\label{eq:kLbench}
\end{equation}
This is in the large-angle CMB range. The suppression at $k\ll k_{\rm L}$ reflects the absence of an arbitrarily long prehistory in which those modes could approach the usual asymptotic Bunch--Davies solution, while the oscillations are interference between the positive- and negative-frequency components created at matching. Larger $s_i$ produces more total inflation and shifts $k_{\rm L}$ below the present horizon, making this signal unobservable. The mechanism is therefore testable but does not force the feature to be visible without a measure for $s_i$.

\begin{figure}[t]
\centering
\includegraphics[width=\columnwidth]{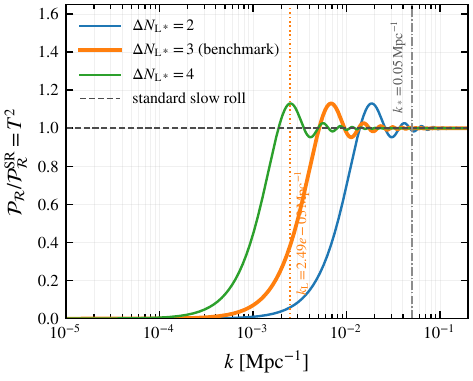}
\caption{Finite-onset transfer function $T^2(k/k_{\rm L})$ for two, three, and four e-folds between release and pivot exit. The thick curve is the benchmark $\Delta N_{\rm L*}=3$, with $k_{\rm L}=2.49\times10^{-3}\,{\rm Mpc}^{-1}$. The observable imprint is a $k^2$ suppression at $k\ll k_{\rm L}$ followed by decaying oscillations and recovery of the standard slow-roll spectrum at $k\gg k_{\rm L}$.}
\label{fig:finite-onset}
\end{figure}

Figure~\ref{fig:finite-onset} isolates the mechanism-specific two-point imprint generated by imposing the positive-frequency state at a finite onset time. Changing the interval $\Delta N_{\rm L*}$ shifts the characteristic scale $k_{\rm L}=k_*\e^{-\Delta N_{\rm L*}}$ while preserving the same low-$k$ suppression, oscillatory transition, and ultraviolet recovery. For the benchmark $\Delta N_{\rm L*}=3$, the feature lies in the large-angle CMB range; hence it is testable through a dedicated low-multipole likelihood, whereas modes with $k\gg k_{\rm L}$ remain observationally indistinguishable from ordinary slow roll.

\begin{figure*}[!t]
\centering
\includegraphics[width=0.94\textwidth]{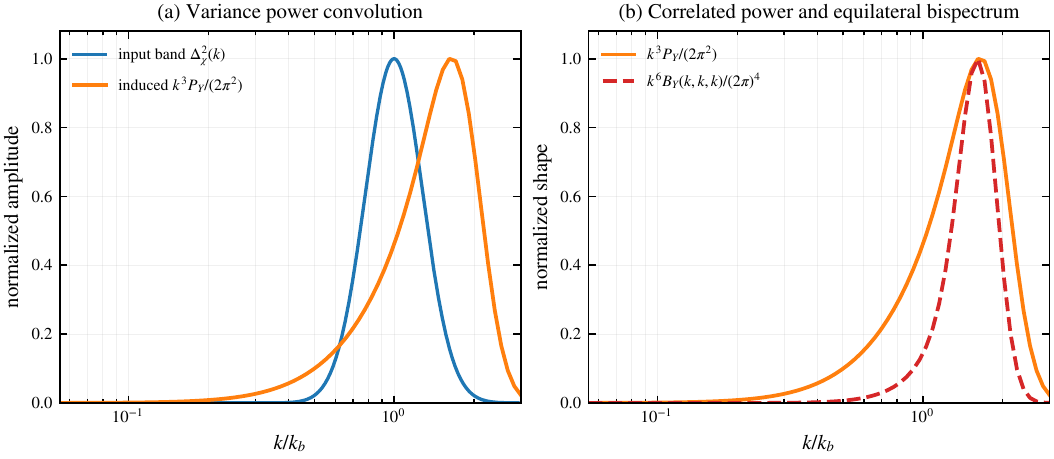}
\caption{Variance-convolution signature for the illustrative narrow lognormal band in Eq.~\eqref{eq:illustrative-band}. Left: the input dimensionless preparation spectrum and the induced variance power $k^3P_Y/(2\pi^2)$. Right: the normalized induced power and equilateral bispectrum shape $k^6B_Y(k,k,k)/(2\pi)^4$. Both observables occupy the same characteristic band, implementing the correlated test. The curves are normalized to unit maximum because their absolute amplitudes require nonlinear evolution of the preparation spectrum and a dedicated CMB likelihood.}
\label{fig:variance-convolution}
\end{figure*}

\subsection{Variance-modulated release}

\label{sec:NG}

Define the local end of preparation by a fixed small threshold
\begin{equation}
\sigma^2(\bm x,N_{\rm L}(\bm x))=\sigma_f^2.
\label{eq:release-surface}
\end{equation}
Write
\begin{equation}
\sigma^2(\bm x,N)=\bar\sigma^2(N)+\delta\sigma^2(\bm x,N).
\end{equation}
Because $\bar\sigma^2\propto\e^{-2N}$,
\begin{equation}
\frac{\dd\bar\sigma^2}{\dd N}=-2\bar\sigma^2.
\end{equation}
Perturbing Eq.~\eqref{eq:release-surface} gives
\begin{equation}
0=\delta\sigma^2-2\sigma_f^2\delta N_{\rm L}.
\end{equation}
Therefore the release contribution to the curvature perturbation is
\begin{equation}
\zeta_{\rm L}=\delta N_{\rm L}
=\frac{\delta\sigma^2}{2\sigma_f^2}.
\label{eq:zetaL}
\end{equation}
This is the separate-universe time-delay mechanism~\cite{LythRodriguez2005,AgarwalHolmanTolleyLin2013}. A region with a slightly larger local variance reaches the release threshold later, accumulates a different local expansion, and acquires a curvature perturbation even though the microscopic preparation field has zero mean. Because the clock is $\chi^2$ rather than $\chi$, the leading release perturbation is intrinsically non-Gaussian.
For an underlying Gaussian preparation field,
\begin{equation}
\delta\sigma^2(\bm x)=\chi^2(\bm x)-\avg{\chi^2},
\end{equation}
so $\zeta_{\rm L}$ is a shifted chi-square field.

Let
\begin{equation}
\avg{\chi_{\bm k}\chi_{\bm k'}}
=(2\pi)^3\delta^{(3)}(\bm k+\bm k')P_\chi(k).
\end{equation}
For
\begin{equation}
Y_{\bm k}=\int\frac{\dd^3p}{(2\pi)^3}
\chi_{\bm p}\chi_{\bm k-\bm p},
\qquad \bm k\ne0,
\end{equation}
Wick's theorem gives
\begin{align}
P_Y(k)&=2\int\frac{\dd^3p}{(2\pi)^3}
P_\chi(p)P_\chi(\abs{\bm k-\bm p}),
\label{eq:PY}\\
B_Y(k_1,k_2,k_3)
&=8\int\frac{\dd^3p}{(2\pi)^3}P_\chi(p)
P_\chi(\abs{\bm p-\bm k_1})\nonumber\\
&\hspace{1.5cm}\times P_\chi(\abs{\bm p+\bm k_2}).
\label{eq:BY}
\end{align}
Equation \eqref{eq:zetaL} implies
\begin{align}
P_{\zeta_{\rm L}}(k)&=\frac{P_Y(k)}{4\sigma_f^4},
\label{eq:PzL}\\
B_{\zeta_{\rm L}}(k_1,k_2,k_3)&=\frac{B_Y(k_1,k_2,k_3)}{8\sigma_f^6}.
\label{eq:BzL}
\end{align}
A correlated observational test is therefore
power-spectrum feature at $k_{\rm L}$, variance-convolution bispectrum at the same scale.
The amplitude is not fixed until the preparation-band spectrum is specified and evolved nonlinearly. The convolution structure nevertheless fixes an important qualitative correlation: the same band that determines the local release time determines both the feature in the power spectrum and the bispectrum support.

To display this fixed convolution morphology, consider the explicit illustrative band
\begin{align}
\Delta_\chi^2(k)
&=\exp\!\left[-\frac{\ln^2(k/k_b)}{2\Delta_b^2}\right],\nonumber\\
P_\chi(k)
&=\frac{2\pi^2}{k^3}\Delta_\chi^2(k),
\qquad \Delta_b=0.25.
\label{eq:illustrative-band}
\end{align}
Substitution into Eqs.~\eqref{eq:PY}--\eqref{eq:BY} gives the normalized templates in Fig.~\ref{fig:variance-convolution}. Their normalization is deliberately removed; only the correlated scale dependence is displayed.

Figure~\ref{fig:variance-convolution} displays the correlated higher-order imprint of the release mechanism. Because the local release clock is proportional to $\chi^2$, the same preparation band that sources the additional curvature power also fixes the support of the variance-convolution bispectrum. The physically robust prediction is therefore the coincidence of the power-spectrum and bispectrum feature scales, not a universal numerical amplitude. A joint search for both structures around the same $k$ range provides a sharper falsifiability criterion than fitting an isolated suppression or an unconstrained non-Gaussian template.

\section{Conclusions}
\label{sec:conclusion}

In this work, we have developed a minimal mechanism for dynamically preparing suitable initial conditions for exponential plateau inflation within ordinary Einstein gravity and using only the inflaton degree of freedom. The central observation is that the variance of inhomogeneous inflaton excitations modifies the coarse-grained effective potential through exact Gaussian averaging,
thereby generating a variance-dependent Hartree minimum.
Consequently, sufficiently large initial fluctuations can dynamically favor configurations in which the coarse-grained inflaton is displaced toward the inflationary plateau. Within the microcanonical measure adopted here, the available phase-space volume is strongly concentrated around this minimum, while configurations carrying an order-one fraction of the total energy in homogeneous kinetic motion are correspondingly suppressed. The statistical preparation of the inflaton is therefore a consequence of the phase-space measure rather than an independently imposed homogeneous initial condition.

The subsequent cosmological evolution preserves the essential outcome of this preparation mechanism. The inhomogeneous variance and its associated energy density redshift as
\begin{equation}
\sigma^2\propto a^{-2},
\qquad
\rho_\chi\propto a^{-4},
\end{equation}
whereas residual homogeneous kinetic energy is diluted even more rapidly, as $a^{-6}$. Although the instantaneous Hartree minimum moves rapidly back toward the vacuum as the fluctuations redshift, the force acting on a field already displaced onto the plateau remains exponentially suppressed. The resulting analytic bounds therefore ensure that the coarse-grained inflaton does not simply follow the retreating minimum, but instead remains at sufficiently large field values with a sufficiently small velocity for slow-roll inflation to commence. This behavior was quantified through the sufficient condition derived in Eq.~\eqref{eq:sufficient-condition}. For the explicit benchmark considered here, the mechanism guarantees at least $63$ slow-roll e-folds while reducing the homogeneous kinetic-energy fraction below $1.3\times10^{-4}$.

At the level of asymptotic slow-roll observables, the benchmark remains compatible with current CMB-only constraints. Its status becomes less favorable once the most recent combined datasets are included: within baseline $\Lambda$CDM, the CMB-SPA+DESI DR2 inference of the scalar spectral index lies $2.66\sigma$ above the benchmark value. The model should therefore not be interpreted as being in agreement with all presently available cosmological data. This distinction is important because the principal result of the present analysis is the dynamical preparation mechanism itself, rather than an attempt to optimize the plateau potential against the latest parameter constraints.

A potentially more distinctive consequence arises when the onset of slow roll occurs sufficiently close to the horizon exit of observable modes. In this regime, the preparation stage can leave finite-onset signatures in the primordial perturbations. At the two-point level, the characteristic prediction is a suppression of power on sufficiently large scales, accompanied by decaying oscillations around a transition scale $k_{\rm L}$ and recovery of the standard slow-roll spectrum at $k\gg k_{\rm L}$. Spatial modulation of the release time provides a complementary higher-order signature: because the local release condition depends on the variance of the preparation field, the same fluctuation band that contributes additional curvature power also generates a variance-convolution bispectrum. The robust feature is therefore not a universal amplitude, but the correlated localization of the power-spectrum and bispectrum structures around the same characteristic scale. Their absolute amplitudes remain dependent on the initial excitation spectrum and on the nonlinear dynamics of the release process.

Taken together, these results provide a concrete proof of principle that inhomogeneous inflaton fluctuations can act as a dynamical preparation mechanism for plateau inflation without introducing additional fields or modifying Einstein gravity. The mechanism simultaneously supplies a statistical preference for high-field configurations, dynamically suppresses excessive homogeneous kinetic energy, and admits analytic conditions under which a sufficiently long slow-roll phase follows. Its phenomenological relevance is ultimately testable: a near-minimal inflationary history may retain correlated two- and three-point signatures of the preparation stage, whereas a longer subsequent inflationary phase shifts these features beyond the observable window. Determining the absolute size of these signatures requires specifying the initial excitation spectrum and following its nonlinear evolution through the release transition, and constitutes the natural next step toward a quantitative confrontation of the mechanism with CMB data.

\end{document}